\pdfoutput=1
\documentclass[aps,prd,twocolumn,preprintnumbers,floatfix,superscriptaddress,nofootinbib]{revtex4-2}

\usepackage{graphicx}
\usepackage{amsmath, amssymb, mathrsfs}
\usepackage{color}
\usepackage{bm}
\usepackage{nccmath, mathtools}
\usepackage{braket} 
\usepackage[T1]{fontenc}
\usepackage[utf8]{inputenc}
\usepackage{microtype}
\usepackage[normalem]{ulem}
\usepackage[dvipsnames]{xcolor}
\usepackage[colorlinks,urlcolor=NavyBlue,citecolor=NavyBlue,linkcolor=NavyBlue,pdfusetitle]{hyperref}
\usepackage[all]{hypcap}
\usepackage{orcidlink}

\newcommand{\pd}{\partial}

\renewcommand{\Im}{\text{Im}}

\graphicspath{{figs/}}

\newcommand{\codename}[1]{{\fontfamily{qcr}\selectfont #1}}

\begin{document}

\title{Approximate helical symmetry in compact binaries}

\author{Aniket Khairnar\,\orcidlink{0000-0001-5138-572X}}
\email{akhairna@go.olemiss.edu}
\affiliation{Department of Physics and Astronomy,
    University of Mississippi, University, MS 38677, USA}

\author{Leo C.\ Stein\,\orcidlink{0000-0001-7559-9597}}
\affiliation{Department of Physics and Astronomy,
    University of Mississippi, University, MS 38677, USA}

\author{Michael Boyle\,\orcidlink{0000-0002-5075-5116}}
\affiliation{Cornell Center for Astrophysics and Planetary Science, Cornell University, Ithaca, New York 14853, USA}

\hypersetup{pdfauthor={Khairnar, Stein, and Boyle}}

\begin{abstract}
    The inspiral of a circular, non-precessing binary exhibits an
    approximate helical symmetry.  The effects of eccentricity,
    precession, and radiation reaction break the exact symmetry.  We
    estimate the failure of this symmetry using the flux of the Bondi–van der
    Burg–Metzner–Sachs   (BMS) charge corresponding to helical symmetry carried
    away by gravitational waves. We analytically compute the helical flux for
    binaries 
    moving on eccentric orbits and quasi-circular orbits without precession
    using post-Newtonian theory. The helical flux is non-vanishing at the 0PN
    order for eccentric orbits as expected. We analytically predict the helical
    flux to be at a relative 5PN order for quasi-circular non-precessing
    binaries. This prediction is compared with 113 quasi-circular non-precessing
    numerical relativity waveforms from the SXS catalog. We find good agreement
    between analytical and numerical results for quasi-circular non-precessing
    binaries establishing that helical symmetry starts to break at 5PN
    for these binaries.
\end{abstract}

\maketitle

\section{Introduction}
\label{sec:introduction}
Einstein's field equations of general relativity have a non-linear structure. This
leads to significant difficulties in solving them analytically for
arbitrary gravitational sources.  Approximate analytical techniques like
post-Newtonian (PN) theory and black hole perturbation theory (BHPT) are used to
solve them for the cases of either weak fields and slow motions, or
a sufficiently small perturbation to a black hole spacetime.
But these methods are invalid close to the merger of a black hole
binary, when GR is highly dynamical and non-linear.
Only numerical relativity (NR) can solve Einstein's equations
in this strong field regime.  However since NR is expensive, we try
to exploit analytical results whenever possible, and using
symmetries---even approximate symmetries---is indispensable.  In this
work, we quantify an under-appreciated approximate symmetry exhibited by
compact binary systems.

When a compact binary has a closed circular
orbit with no precession, then the spacetime exhibits an exact
helical Killing vector. This concept of helical symmetry has been used
to impose quasi-stationarity in the initial data of binary black
holes~\cite{Gourgoulhon:2001ec, Grandclement:2001ed}, and to study
orbital dynamics in PN theory~\cite{Blanchet:2023sbv,
Bernard:2017ktp, Blanchet:2024mnz, Faye:2012we, Arun:2007rg,
Blanchet:1996wx}. In recent works, this symmetry has been used to establish laws
of binary black hole mechanics for non-spinning~\cite{LeTiec:2011ab} and
spinning particles~\cite{Ramond:2022ctc} in studies of the
gravitational self-force. However in a real inspiralling binary, the
helical symmetry is only approximate, restricting its applicability.

For an inspiralling quasi-circular binary, or one with a precessing
orbital plane, or one with
eccentric orbits, this symmetry is broken. The effects of eccentricity,
precession, and radiation reaction break the symmetry in decreasing order of
importance. We say such spacetimes exhibit an approximate helical
symmetry based on the timescales over which these effects operate. The
eccentricity is certainly the dominant effect because it operates on an orbital
timescale, completely breaking the symmetry.  Precession of the orbital
plane operates on timescale longer than the orbital timescale, while radiation
reaction operates on the longest.  Therefore radiation reaction is the weakest
effect to break the helical symmetry.   

We want to analytically compute the effect of approximate symmetry breaking
using PN theory.
However we are limited by today's PN results not going to sufficiently
high order.
Thus we quantify the approximate helical symmetry by comparing it with
the SXS catalog of NR simulations~\cite{Boyle:2019kee}, supporting our
analytical understanding.
We use the
initial orbital parameters of the binary system, like eccentricity and
dimensionless spin vectors, to sort the simulations into categories of eccentric
non-precessing, quasi-circular precessing, and quasi-circular non-precessing
systems.  In this way, we isolate the various effects that break the helical
symmetry.

Specifically, to quantify the symmetry, we will apply Noether's
theorem~\cite{Noether:1918zz} to compact binaries' gravitational
waves, which encode the approximate helical symmetry.
From Noether's theorem, the charge corresponding to a continuous symmetry
would be conserved, whereas a broken symmetry implies a non-vanishing
flux of the associated charge at future null infinity, $\mathscr{I}^{+}$.
Thus gravitational waves carry away a flux of the helical charge. The behavior of
the helical flux helps us to establish its dependence on the source parameters
like mass ratio, frequency, and eccentricity.
Our analysis underscores that one cannot always ignore the breaking of
the helical symmetry. As suggested by~\cite{Apostolatos:1993nu}, this implies
that quasi-circular orbits remain quasi-circular under radiation reaction, to a
very high PN order.

We begin with Section~\ref{sec:helical symmetry} by introducing the concept of
helical symmetry and the corresponding helical flux mathematically. We present
the non-vanishing helical fluxes for the three effects appearing at different PN
orders. Then we provide analytical results for the effect of eccentricity and
radiation reaction on breaking the symmetry in
Section~\ref{sec:analyticalresults}. We find the analytical result of
helical flux for an eccentric non-precessing binary to be at a relative 0PN to
energy flux. Then we proceed to the analytical computation of helical flux for a
quasi-circular non-precessing binary that is relatively complex. We find it to
be at a relative 5PN order to the energy flux that matches with our numerical
results presented in Section~\ref{sec:numericalresults}. In
Section~\ref{sec:disc-concl}, we conclude with our discussion on the choice of
angular velocity in numerical analysis, the modulations observed in the helical
flux, and the prospects of our work in the future.

\section{Helical symmetry}
\label{sec:helical symmetry}
\begin{figure}
    \centering
    \includegraphics[height=6cm,trim=0 40 0 0]{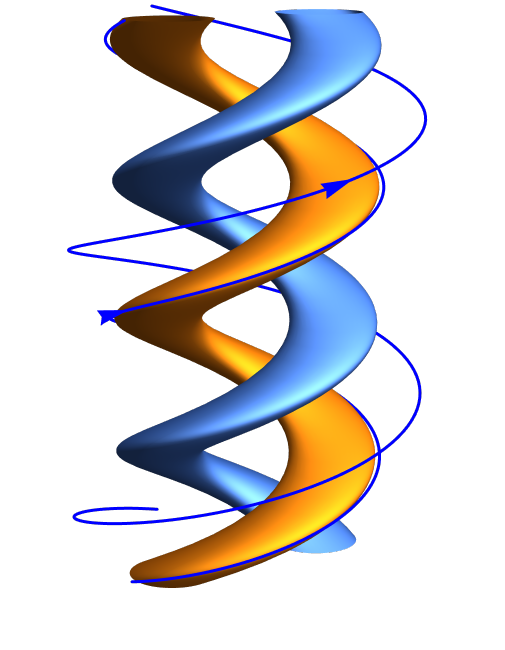}
    \caption{%
      \label{fig:helical-diagram}%
      Schematic representation of a binary black hole spacetime with a
      helical symmetry.}
\end{figure}

Helically symmetric spacetimes have a global symmetry generated by
a helical Killing vector field. Mathematically it can be represented
as a combination of time translation and rotation as
\begin{equation}
    K^{\alpha} \pd_{\alpha} = \pd_t + \Omega \, \pd_{\phi},
\end{equation}
where $\pd_t$ is the generator of time translations and $\pd_{\phi}$
is the generator of rotation about a symmetry axis; in an asymptotic
region, we can say the rotation is about a vector $\hat{L}$.  Here $\Omega$ is
the asymptotic frequency of the rotation measured by observers at
future null infinity.
For this symmetry to be exact, $\Omega$ must simply be a constant.
A spacetime exhibiting this symmetry is represented in
Fig.~\ref{fig:helical-diagram}.

Our systems of interest are slowly-evolving compact binaries,
inspiralling due to radiation reaction.  Here $\Omega$ is
approximately half the frequency of the dominant $(2,2)$ mode
radiation (and is related to the orbital frequency of the binary
measured by local observers as shown in \cite{Blanchet:2024mnz}).

The effects of eccentricity, precession, and radiation reaction break
the exact symmetry to only an approximate symmetry, which we want to quantify.
As per Noether's theorem, the charge corresponding to
a global symmetry would be conserved.  Therefore the flux of this
charge would be identically zero for an unbroken symmetry.

For asymptotically flat spacetimes, the asymptotic symmetry group at
$\mathscr{I}^{+}$ is not the Poincaré group, but an enlarged group
called the Bondi--van~der~Burg--Metzner--Sachs (BMS)
group~\cite{Bondi:1962px,Sachs:1962zza,Sachs:1962wk}.
The BMS group is a semidirect product between the Poincaré group and
supertranslations, which are interpreted as direction-dependent
translations.  Supertranslations form an infinite-dimensional Abelian
subgroup.  Every generator in the BMS algebra, given by a vector field $\xi$,
gives rise to an associated flux at
$\mathscr{I}^{+}$~\cite{Wald:1999wa,Flanagan:2015pxa}
\begin{equation}
    F_{\xi} =-\frac{1}{32 \pi} \int d^2 x \, N^{AB} \delta_{\xi} C_{AB}, \label{eq:BMS flux}
\end{equation}
where $C_{AB}$ is the shear and $N_{AB}$ is the Bondi
news. Here $\delta_{\xi} C_{AB}$ is the change in the shear resulting from
$g_{ab} \to g_{ab} + \mathcal{L}_{\xi} g_{ab} $
\cite{Flanagan:2015pxa}.

When $\Omega$ is time-dependent, $K^{\alpha}\pd_{\alpha}$ is not in the BMS algebra.
However, when $\Omega$ is constant, we have a helical symmetry and
$K^{\alpha}\pd_{\alpha}$ is a BMS generator (specifically, a Poincaré generator).
The helical vector field can be expressed in Bondi coordinates as
\begin{equation}
  \xi = K^{\alpha}\pd_{\alpha} = \pd_u + (\vec{\Omega} \times \vec{x}) \cdot \vec{\pd}
  \,
  . \label{eq:helical vector field}
\end{equation}
Using the linearity of Lie derivatives it can be shown that
\begin{equation}
  F_{\xi} =  \dot{E}  - \boldsymbol{\Omega} \cdot \dot{\textbf{L}}
  \,, 
\end{equation} 
where $\dot{E}$ is energy flux, $\dot{\textbf{L}}$ is angular momentum
flux, and $\xi$ is the helical vector field as defined in
Eq.~\eqref{eq:helical vector field}.

\begin{figure}
    \centering
    \includegraphics{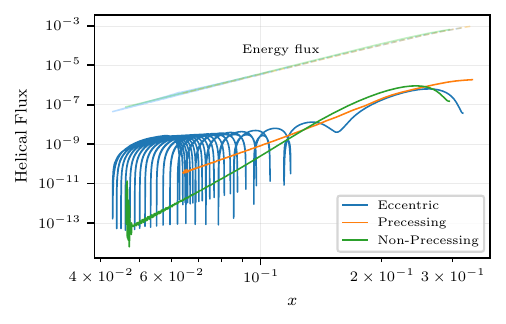}
    \caption{%
      \label{fig:ecc-prec-qc-flux}%
      Helical flux computed from an eccentric (green, SXS:BBH:2599,
      $e = 0.21$), precessing (orange, SXS:BBH:2450,
      $\chi_{1,2}^{\perp} \approx 0.5$) and non-precessing
      quasi-circular (blue, SXS:BBH:3912) SXS simulation.  These
      fluxes are plotted as a function of the post-Newtonian parameter
      $x$, Eq.~\eqref{eq:x-PN}.  For reference, the energy fluxes for
      the three systems are plotted (same colors but dashed) on the
      same axes.}
\end{figure}

A real binary has a time-dependent frequency, so we promote
$\boldsymbol{\Omega}$ to $\boldsymbol{\Omega}(u)$, and will quantify the
failure of satisfying the exact symmetry by computing
\begin{align}
  F_{\xi} =  \dot{E}  - \boldsymbol{\Omega}(u) \cdot \dot{\textbf{L}}
  \,. \label{eq: helical flux}
\end{align}
Eccentricity, precession, and radiation reaction will all produce
non-vanishing helical fluxes at $\mathscr{I}^{+}$ at different PN
orders.
Figure~\ref{fig:ecc-prec-qc-flux} demonstrates the lack of
cancellation in an eccentric system (blue), followed by cancellations
at higher PN orders in a precessing system (orange) and finally a
non-precessing quasi-circular system (green).

\section{Analytical results}
\label{sec:analyticalresults}

Let us analytically compute the
helical flux for eccentric binary systems and quasi-circular, non-precessing
systems.  The inspiral of a binary evolves gradually due to the emission of
gravitational radiation.  We employ the post-Newtonian multipolar post-Minkowskian
(PN-MPM) formalism as discussed in~\cite{Blanchet:2024mnz} to study such
slow-moving and weak field sources.

\subsection{Eccentricity}
The helical symmetry is intuitively absent in eccentric binaries to leading
order in perturbation theory. A leading order PN-MPM computation of the helical
flux would suffice to obtain the non-vanishing helical flux. We follow the
approach outlined in~\cite{Blanchet:2024mnz, Loutrel:2016cdw} using the
quasi-Keplerian parametrization of the motion of the binary. We review the steps
necessary to get the helical flux for an eccentric binary.

The binary system is described in the following way. We have two bodies with masses $m_1$ and $m_2$ moving on an elliptical orbit with
eccentricity $e$ and semimajor axis $a$.  In the reduced one-body description, a
body of reduced mass $\mu=m_{1}m_{2}/(m_{1}+m_{2})$ is orbiting around
a central body of mass $m= m_1 + m_2$, that is situated at one of the
foci of the ellipse.  We primarily use the symmetric mass ratio
$\nu=\mu/m$ instead of $\mu$.  The motion is planar and
perpendicular to the direction of orbital angular momentum $L$. The
orbital separation is $r=|\textbf{r}|$ and the relative velocity is
$\textbf{v}$. The radius $r$ and orbital phase or true anomaly $\phi$ can be
expressed in terms of the eccentric anomaly $u$ as
\begin{align}
    r &= a ( 1 - e \cos u ) + \mathcal{O} \left(\frac{1}{c^8} \right), \\
    \phi - \phi_p &= 2 K \arctan \left[ \left(\frac{1+e}{1-e} \right)^{1/2} \tan \frac{u}{2} \right] + \mathcal{O} \left( \frac{1}{c^4} \right),
\end{align}
where $\phi_p$ is the phase at the pericenter, and $K$ is the precession of the
pericenter per orbit.
The eccentric anomaly is related to the mean anomaly $l$ by Kepler's
equation, which to leading order is
\begin{align}
  l = n (t - t_p) = u - e \sin u + \mathcal{O} \left(\frac{1}{c^4} \right)
  \,,
\end{align}
where $n$ is the mean motion, and $t_p$ is the time of pericenter passage.
These orbital elements can be expressed in terms of the dimensionless
energy $\varepsilon$ and dimensionless angular momentum $j$,
defined in terms of the energy $E$ and specific angular momentum $h=J/Gm$ as
\begin{align}
  \varepsilon = -\frac{2E}{\mu c^2},   \quad
  j = -\frac{2E h^2}{\mu^3}.
\end{align}
The expansion parameter $x$ in PN theory for eccentric binaries is related to the
orbital frequency $\Omega$ and the dimensionless parameters as
\begin{subequations}
  \label{eq:x-PN}
\begin{align}
  x &= \left( \frac{G \, m  \, \Omega}{c^3} \right)^{2/3} \,, \\
  x &= \varepsilon \left\{ 1 + \varepsilon \left[ - \frac{5}{4} + \frac{1}{12} \nu + \frac{2}{j} \right] + \mathcal{O} (\varepsilon^2) \right\}
  \,.
\end{align}
\end{subequations}
The orbital elements defined earlier can be expanded in terms of
$x$ using intermediate relations with $\varepsilon$
given in \cite{Blanchet:2024mnz}
\begin{subequations}
  \begin{align}
    a &= \frac{Gm}{x c^2} \left\{1 + \mathcal{O}\left(\frac{1}{c^2}\right)\right\}, \\
    n &= \frac{x^{3/2} c^3}{G m} \left\{1 + \mathcal{O}\left(\frac{1}{c^2}\right)\right\}, \\
    K &= 1 + \mathcal{O}\left(\frac{1}{c^2}\right).
  \end{align}
\end{subequations}

The contribution from the source quadrupole moment will be sufficient
to obtain the helical flux to the leading order. The quadrupole moment at Newtonian order is given by 
\begin{align}
    \text{I}_{ij} = \nu \, m \, x_{\langle i} x_{j \rangle}.
\end{align}
It is straightforward to compute the helical flux using Eq.~\eqref{eq: helical
flux}, computing energy and angular momentum flux following~\cite{Blanchet:2024mnz}. We provide the instantaneous results for the
fluxes instead of orbit averaged expressions,
\begin{align}
    \dot{E} &= \frac{32 c^5 \nu ^2 x^5}{5 G (1-e \cos u)^6} \left(1-\frac{1}{24} e^2 \cos 2 u-\frac{23 e^2}{24}\right),\\
    \dot{L} &= \frac{32 c^2 m \nu ^2 x^{7/2} \sqrt{1-e^2}}{5 (1-e \cos u)^5} \bigg(1 - \frac{1}{2} e \cos u \nonumber \\ 
    & \qquad \qquad -\frac{5 e^2}{8}
    + \frac{1}{8} e^2 \cos 2 u  \bigg).
\end{align}
\begin{widetext}
The instantaneous helical flux for the eccentric binary is
\begin{subequations}
  \begin{align}
    F_{\xi} &= \frac{32c^5 \nu^2 x^{5}}{5G (1- e \cos u)^6} \left\{ 1 -  \frac{23 e^2}{24} -\frac{e^{2}\cos 2u}{24}
              - \sqrt{1-e^2}
              (1-e\cos u)
              \left(
              1 - \frac{1}{2} e \cos u -\frac{5 e^2}{8} + \frac{1}{8} e^2 \cos 2 u \right)      \right\} \,,\\
    &= \frac{48 c^5 \nu ^2 x^5 e \cos u}{5 G}+\mathcal{O}\left(e^2\right)  .
    \end{align}
\end{subequations}
We present the orbit averaged expression for this flux as well. The expression
matches with the averaged helical flux derived in \cite{LeTiec:2015kgg} for eccentric orbits
\begin{subequations}
  \begin{align}
    \langle F_{\xi} \rangle &= \frac{32}{5G}  c^5 \nu^2 x^{5} \left\{\frac{1}{(1-e^2)^{7/2}} \left( 1+ \frac{73}{24} e^2 +  \frac{37}{96} e^4 \right) - \frac{1}{(1-e^2)^2} \left( 1 + \frac{7}{8}e^2 \right)   \right\} \,,\\
    &= \frac{352 c^5 \nu^2 x^5 e^2}{15 G} +\mathcal{O}\left(e^4\right)  .
  \end{align}
\end{subequations}
\end{widetext}
The instantaneous and the averaged flux are non-linear in eccentricity and
start at relative 0PN order with respect to the energy flux, as expected. In the
limit of small eccentricity, they yield a simpler dependence on eccentricity.

\subsection{Quasicircular, nonprecessing}
Recent works have pushed PN results to 4.5PN order beyond the leading
quadrupole radiation~\cite{Blanchet:2023sbv,Marchand:2016vox}. This limits our
helical flux calculation to 4.5PN accuracy.

To peek beyond the 4.5PN order, we estimate the leading order
contribution to the helical flux from the source's
$l$\textsuperscript{th} mass or current multipole moment. This
approach gives a general understanding of the contribution to the
helical flux from each multipole moment. We compare this prediction
with the helical flux computed using numerical relativity
simulations of binary black hole mergers.
\subsubsection{Post-Newtonian helical flux}
The helical flux requires the computation of energy and angular
momentum flux as shown in Eq.~\eqref{eq: helical flux}. The energy flux
has been computed in the literature up to 4.5PN order. This computation begins
with constructing the metric perturbation in a Bondi-type coordinate system $(u,
\mathbf{X})$ 
\begin{align}\label{eq:hTT}
h_{ab}^\text{TT} &= \frac{4G}{c^2R} \perp_{abij}(\bm{N})\sum_{\ell \geqslant 2}\frac{1}{c^\ell\,\ell!}\bigg[N_{L-2}\,\text{U}_{ijL-2}(u) \\
& - \frac{2\ell}{c(\ell+1)}N_{kL-2}\epsilon_{kl(i}\text{V}_{j)lL-2}(u)\bigg] + \mathcal{O}\left(\frac{1}{R^2}\right), \, \nonumber
\end{align}
where $N^i = X^i/R$ and $\perp_{abij}$ is the projection
tensor. The metric perturbation is expressed in terms of radiative
mass-type $\text{U}_L$ and current-type $\text{V}_L$ multipole moments in a
transverse-traceless gauge. It is straightforward to obtain the
multipole expansion of the energy flux using Eq.~\eqref{eq:BMS flux}, and is
expressed as
\begin{align}
\dot{E} =  \sum_{l = 2}^{\infty} \frac{G}{c^{2l+1}} \left[ a_l \, \text{U}_{L}^{(1)} \text{U}_{L}^{(1)}  + \frac{b_l}{c^2} \, \text{V}_{L}^{(1)} \text{V}_{L}^{(1)}  \right], \label{eq:Energymultipoleexpansion}
\end{align}
where $a_l$ and $b_l$ are derived in~\cite{Thorne:1980ru}, and the
superscript numbers in parentheses denote retarded time derivatives,
$\text{U}_{L}^{(j)} \equiv \frac{d^{j}}{du^{j}}\text{U}_{L}$. We
account for the non-linear effects in gravitational wave propagation
by relating the radiative multipole moments to the source
$(\text{I}_L, \text{J}_L)$ that describe the PN source. We achieve
this using the intermediate canonical multipole moments
$(\text{M}_L, \text{S}_L)$. The relation between radiative and
canonical multipole moments contain instantaneous and hereditary type
of terms up to a given PN order. Many of these relations have been
derived in references~\cite{Faye:2012we, Faye:2014fra,
  Marchand:2016vox, Marchand:2020fpt, Larrouturou:2021dma,
  Larrouturou:2021gqo, Henry:2021cek, Blanchet:2022vsm, Trestini:2022tot}.  At the stage of deriving the PN series for
source multipole moments $(\text{I}_L, \text{J}_L)$, the procedure is
specialized from generic orbits to quasi-circular orbits in the
center-of-mass frame. Then we use the equations of motion for
particles moving on these orbits for the computation of fluxes. As
shown in~\cite{Blanchet:2024mnz}, this procedure can be used to
compute any Poincaré flux.

The recent 4.5PN-accurate expression for $\dot{E}$ appears in
Eq.~(6.11) of~\cite{Blanchet:2023sbv}, which they call $\mathcal{F}$. Here, we
reproduce this existing computation and perform a similar computation for the
angular momentum flux. The angular momentum flux can be expressed in terms of
the radiative multipole moments $\text{U}_{L}$ and $\text{V}_{L}$ as
\begin{align}
\dot{L}_i =  \epsilon_{i a b} \sum_{l = 2}^{\infty} \frac{G}{c^{2l+1}} \left[ a_l' \, \text{U}_{a L-1} \text{U}_{b L-1}^{(1)}  + \frac{b_l'}{c^2} \, \text{V}_{a L-1} \text{V}_{b L-1}^{(1)}  \right], \label{eq:angfluxradmomexpansion}
\end{align}
where the coefficients $a_l'$ and $b_l'$ are given
in~\cite{Blanchet:2024mnz}.  The 4.5PN-accurate expression for
$\dot{L}$ has not appeared in the literature, as far as we are aware,
though it was \emph{expected} to agree with $\dot{E}/\Omega$.  We have
explicitly computed $\dot{L}$, accurate to 4.5PN order, using the
equations of motion and the series expansions for $\text{U}_{L}$ and
$\text{V}_{L}$ presented in \cite{Blanchet:2023sbv,Blanchet:2024mnz},
using the \codename{xAct/xTensor} suite for
\codename{Mathematica}~\cite{JMM:xAct, MARTINGARCIA2008597}.  The
magnitude of the angular momentum flux is
\begin{widetext}
\begin{align}\label{eq:AngMomFlux}
\dot{L} ={}& \frac{32}{5} c^2 m \nu^2 x^{7/2}\biggl\{
1
+ \biggl[-\tfrac{1247}{336} - \tfrac{35}{12}\nu \biggr] x
+ 4\pi x^{3/2}
\nonumber + \biggl[-\tfrac{44711}{9072} +\tfrac{9271}{504}\nu + \tfrac{65}{18} \nu^2 \biggr] x^2
+ \biggl[-\tfrac{8191}{672}-\tfrac{583}{24}\nu \biggr]\pi x^{5/2}
\nonumber\\
&
+ \biggl[\tfrac{6643739519}{69854400}+ \tfrac{16}{3}\pi^2-\tfrac{1712}{105}\gamma_\text{E} - \tfrac{856}{105} \ln (16\,x)
 + \Bigl(-\tfrac{134543}{7776} + \tfrac{41}{48}\pi^2 \Bigr)\nu
- \tfrac{94403}{3024}\nu^2
- \tfrac{775}{324}\nu^3 \biggr] x^3
\nonumber \\
&
+ \biggl[-\tfrac{16285}{504} + \tfrac{214745}{1728}\nu +\tfrac{193385}{3024}\nu^2 \biggr]\pi x^{7/2} + \biggl[ -\tfrac{323105549467}{3178375200} + \tfrac{232597}{4410}\gamma_\text{E} - \tfrac{1369}{126} \pi^2 + \tfrac{39931}{294}\ln 2 - \tfrac{47385}{1568}\ln 3 + \tfrac{232597}{8820}\ln x
\nonumber \\
&
+ \Bigl( -\tfrac{1452202403629}{1466942400} + \tfrac{41478}{245}\gamma_\text{E} - \tfrac{267127}{4608}\pi^2 + \tfrac{479062}{2205}\ln 2 + \tfrac{47385}{392}\ln 3  + \tfrac{20739}{245} \ln x \Bigr)\nu + \Bigl( \tfrac{1607125}{6804} - \tfrac{3157}{384}\pi^2 \Bigr)\nu^2 
\nonumber \\
&
+ \tfrac{6875}{504}\nu^3 + \tfrac{5}{6}\nu^4 \biggr] x^4 + \biggl[ \tfrac{265978667519}{745113600} - \tfrac{6848}{105}\gamma_\text{E} - \tfrac{3424}{105} \ln (16 \,x) + \Bigl( \tfrac{2062241}{22176} + \tfrac{41}{12}\pi^2 \Bigr)\nu - \tfrac{133112905}{290304}\nu^2 - \tfrac{3719141}{38016}\nu^3 \biggr] \pi x^{9/2} 
\nonumber \\
&
+ \mathcal{O}(x^5) \biggr\}\,.
\end{align}
\end{widetext}
We outline the procedure to compute the hereditary integrals involved in this
computation in the appendix. Using these flux expressions, we compute the
helical flux as expressed in Eq.~\eqref{eq: helical flux},
\begin{align}
F_{\xi} = 0 + \mathcal{O}(\mathnormal{x}^{10}).
\end{align}
This verifies the expectation that the helical flux vanishes up to
4.5PN order.  In the next subsection, we calculate that the first
non-vanishing PN order of the helical flux is at a relative 5PN order.

\subsubsection{Abstract computation of helical flux}
\label{sec:abstr-comp-helic}
Given the limitations of the post-Newtonian results, we perform a
non-rigorous computation to estimate the PN order for the leading
order term in helical flux relative to energy flux. Similar to
Eq.~\eqref{eq:angfluxradmomexpansion}, the helical flux can be
expressed in terms of the radiative multipole moments through a simple
computation
\begin{align}
    F_{\xi} = \sum_{l = 2}^{\infty} & \frac{G}{c^{2l+1}} \Bigg\{ a_l \left( \text{U}_{L}^{(1)} \text{U}_{L}^{(1)} - \Omega^i \, \epsilon_{i a b} \, l \,  \text{U}_{a L-1} \text{U}_{b L-1}^{(1)} \right) \nonumber \\
    & + \frac{b_l}{c^2} \left( \text{V}_{L}^{(1)} \text{V}_{L}^{(1)} -\Omega^i \, \epsilon_{i a b} \, l \, \text{V}_{a L-1} \text{V}_{b L-1}^{(1)}  \right) \Bigg\} . \label{eq:Hel-radexp}
\end{align}
The estimate for the leading order term would require the Newtonian
order relations between the radiative multipole moments and the source
multipole moments that are given by
\begin{align}
\text{U}_{L} &=  \text{I}_{L}^{(l)} + \mathcal{O}\left(\frac{1}{c^3}\right) \,, \\
\text{V}_{L} &=  \text{J}_{L}^{(l)}  + \mathcal{O}\left(\frac{1}{c^3}\right) \,.
\end{align}
For circular, non-spinning binaries these
source multipole moments can be related to orbital parameters of the
binary to leading order as
\begin{subequations}
\begin{align}
\text{I}_{L} &=  \mu \sigma_l (\nu) \ x_{L} + \mathcal{O}\left(\frac{1}{c}\right), \\
\text{J}_{L-1} &=  \mu \sigma_l (\nu) \ \epsilon_{a b \langle i_{l-1}} x_{L-2 \rangle a} v_b  + \mathcal{O}\left(\frac{1}{c}\right),
\end{align}
\end{subequations}
where $\mu$ is the reduced mass and $\sigma_l(\nu)$ is a constant
defined in~\cite{Blanchet:2024mnz}.  The computation needs the
equations of motion of the inspiralling circular binary system written
in the form
\begin{subequations}
  \label{eq:equations of motion}
\begin{align}
x_i &= r \ n_i , \\
v_i &= \epsilon_i^{\ j k} \ \Omega_j \ x_k + \frac{\dot{r}}{r} \ x_i ,
\end{align} %
\end{subequations}%
where $x_i$ and $v_i$ are the radius and velocity vectors of the
reduced mass. For this calculation, we will refrain from explicitly
substituting the equations of motion until the end, leaving
derivatives in the form $x_i^{(k_i)}$. Using
the generalized Leibniz rule, the $l$\textsuperscript{th} mass multipole moment can be
expanded at leading order as
\begin{align}
\text{I}_{L}^{(l)} &= \mu \sigma_{l}(\nu) \sum_{k_1 + \cdots k_l = l} {l\choose k_1, k_2, \cdots, k_l} x_{i_1}^{(k_1)} x_{i_2}^{(k_2)} \cdots x_{i_l}^{(k_l)}. \label{eq:Leibniz rule result}
\end{align}

Since the flux is linear in the multipole moments, we can focus on the
contribution from the $l$\textsuperscript{th} mass multipole moment to understand the
general behavior,
\begin{widetext}
\begin{align}
    \mathcal{F}_{\xi} (\text{U}_{L}) &=  \text{U}_{i_1 L-1}^{(1)} \left( \text{U}_{i_1 L-1}^{(1)} - \Omega^i \, \epsilon_{i \, j \, i_1} \, l \,  \text{U}_{j L-1} \right)
                                = \text{I}_{i_1 L-1}^{(l+1)} \left( \text{I}_{i_1 L-1}^{(l+1)} - \Omega^i \, \epsilon_{i \, j \, i_1} \, l \,  \text{I}_{j L-1}^{(l)} \right), \\
  \label{F_I eqn}
    &= \mu^{2}\sigma_{l}^{2}\sum_{\Sigma m_{i}=l} \sum_{\Sigma k_{i}=l} {l\choose m_1, \cdots, m_l} {l\choose k_1, \cdots, k_l}
      l^2 \, x_{i_1}^{(m_1+1)} \left( x_{i_1}^{(k_1+1)} - \Omega^i \, \epsilon_{i \, j \, i_1} \, x_{j}^{(k_1)} \right) \,
      x_{i_2}^{(m_2)} x_{i_2}^{(k_2)}
      \cdots x_{i_l}^{(m_l)} x_{i_l}^{(k_l)},
\end{align}
where we substitute the radiative multipole moment with the source
multipole moment in the second line.
Using the quasicircular
equations of motion in Eq.~\eqref{eq:equations of motion}, it can be
shown that
\begin{align}
   x_{i_1}^{(k_1+1)} &= \epsilon_{i_1}{}^{j k} \, \Omega_j \,x_{k}^{(k_1)}
    + \left( 1 - \frac{3}{2}k_1 \right) \left(\frac{\dot{r}}{r}\right) x_{i_1}^{(k_1)} + \mathcal{O}\left(\Omega x^{(k_{1})}\frac{\dot{r}^{2}}{r^{2}\Omega^{2}} \right)
=
\epsilon_{i_1}{}^{j k} \, \Omega_j \,x_{k}^{(k_1)} \left[ 1 + \mathcal{O}\left(\frac{\dot{r}}{r\Omega}\right) \right]
\,. \label{x eqn}
\end{align}
The leading term of the above result (at 0PN relative order) cancels
with the second term in Eq.~\eqref{F_I eqn}.
This leads to a manifest cancellation by 2.5PN orders, but there is
also a subsequent cancellation by another 2.5PN orders:
\begin{align}
\mathcal{F}_{\xi}(\text{U}_{L}) &= \mu^{2}\sigma_{l}^{2}\sum_{\Sigma m_{i}=l} \sum_{\Sigma k_{i}=l} {l\choose m_1, \ldots, m_l} {l\choose k_1,\ldots, k_l} \left(\frac{\dot{r}}{r} \right)  \left( 1 - \frac{3}{2}k_1 \right)  l^2 \, x_{i_1}^{(m_1+1)} x_{i_1}^{(k_1)} \cdots x_{i_l}^{(m_l)} \, x_{i_l}^{(k_l)}
\left[1 + \mathcal{O}  \left(\frac{\dot{r}}{r\Omega} \right)\right] \nonumber \\
&= 0 +  \mathcal{O}\left( \dot{E}(\text{U}_{L}) \, x^{5} \right)
\,.
\end{align}
\end{widetext}
The further cancellation by 2.5PN orders is due to the properties of a
circular orbit.  The 2.5PN order terms have $(2l+1)$ derivatives
distributed over $(2l)$ separation vectors $x_{i}$, which are
contracted together into several scalar terms.
For a circular orbit in a plane, a term of the form
$x_{i_{j}}^{(m_{j})}x_{i_{j}}^{(k_{j})}$ with an odd difference in
derivatives leads to a contraction between perpendicular vectors in
the plane.
Therefore the only types of
contractions that survive at the leading order are of the form
$x_{i_{j}}^{(m_{j})}x_{i_{j}}^{(k_{j})}$ where the derivatives differ
by an even integer, $m_{j}\equiv k_{j}\text{ (mod 2)}$.  However, with
$(2l+1)$ derivatives spread over $(2l)$ vectors, there will always be
at least one term with an odd difference.
Thus the first correction will be
trivially zero and the first non-zero correction to the helical flux
will appear at $\mathcal{O}(\dot{r}/r)^2$, that is a relative 5PN
order to the energy flux contribution from the radiative multipoles
$\text{U}_L$ or $\text{V}_L$. We can see this computation for the quadrupole
moment as
\begin{align}
  \mathcal{F}_{\xi}(\text{U}_{ij}) &= \frac{212992 c^5}{875 G} \nu^4 x^{10} + \mathcal{O}(x^{21/2}) \, .
\end{align}
Thus the contribution to the helical flux from the quadrupole moment is at a
relative 5PN order from its contribution to the energy flux. Though this method
is not rigorous, it agrees with our numerical results below.

\section{Numerical results}
\label{sec:numericalresults}
We take ideas from existing analytical results to obtain comparable
predictions of the PN order of the helical flux from numerical
simulations. In the PN formalism~\cite{Blanchet:2024mnz}, the
frequency related parameter $x$ is used to express the analytical
results for fluxes and other relevant quantities. The helical flux
would have a similar series in $x$ given by
\begin{equation}
  F_{\xi} = a_0 \, x^{p} (1 + a_1 \, x + a_2 \, x^2 + \dots ),
\end{equation}
where $a_i$ and $p$ are as yet unknown.
The numerical relativity waveforms from quasi-circular binaries give
an advantage when estimating such unknown power law behavior. Since
the effect is anticipated at a high PN order, the leading power can be
obtained through an instantaneous PN order computation
\begin{align}
    \textbf{PN Order} = p &\approx  \frac{\mathrm{d}\log F_{\xi}}{\mathrm{d}\log \mathnormal{x}}, \nonumber \\
   p &\approx \frac{3}{2} \times \frac{\mathrm{d}\log F_{\xi}}{\mathrm{d}\log \Omega}.
\end{align}

We use 113 quasi-circular ($e < 10^{-4}$) non-precessing
$\chi_{i \perp} < 10^{-4}$) NR simulations from the SXS
catalog~\cite{Boyle:2019kee} to perform this analysis.  Numerically,
we compute the flux and angular velocity using the \codename{scri}
python package~\cite{boyle_2024_12585016, Boyle:2013nka}.  We compare
this numerical estimate of the instantaneous PN order with the
prediction from the analytical results presented in the previous
section. This numerical analysis runs into some subtleties.

The first obstacle arises from a lack of understanding of a unique
definition of angular velocity to be obtained from numerical
waveforms. In this work, we choose to work with two definitions of angular
velocity: one that tries to maintain the strain to be a constant in
a co-rotating frame~\cite{Boyle:2013nka}, and another obtained from
the phase of the $h_{22}$ mode. We compare the results of these two
approaches below.

The second set of issues is associated with the numerical attributes
of the gravitational waveforms. The high degree of cancellation in the
helical flux is due to the circular nature of the orbit. However, the
numerical waveforms aren't produced from perfectly circular
simulations.  The helical flux is modulated by the presence of
eccentricity.  We handle these modulations by smoothing the flux with a
Gaussian kernel.  The next subsections present the mathematical
details of the choice of angular velocity and the smoothing process.

\subsubsection{Angular velocity by minimal rotation}
A commonly-used definition for quasi-circular binaries was proposed by
Boyle in~\cite{Boyle:2013nka}, inspired by earlier
work~\cite{OShaughnessy:2011pmr}.  Boyle's proposal is based on
finding an instantaneous rotation to keep a waveform as constant as
possible in a corotating frame, by minimizing
\begin{align}
  \Xi(\vec{\Omega}) = \int_{S^{2}} |i \ \vec{\Omega}\cdot\vec{L} \ f+\pd_{t}f|^{2} \ \mathrm{d^2 S} \,, \label{eq:minimal rotation condition}
\end{align}
where the $\vec{L}$ differential operator is the infinitesimal
generator of rotations, $L_{a}=-i \epsilon_{ab}{}^{c}x^{b}\pd_{c}$.
Here the waveform $f(u,\theta,\phi)$ can be e.g.\ the GW strain or the
news.  The optimal $\vec{\Omega}$ is found as
\begin{align}
\label{eq:angularvelocity}
\vec{\Omega}_{\text{rot}}(u)=-\langle\vec{L}\vec{L}\rangle^{-1}\cdot\langle\vec{L}\pd_{t}\rangle,
\end{align}
where the vector and matrix here have components
\begin{subequations}
\begin{align}
\langle\vec{L}\pd_{t}\rangle^{a} & \equiv \sum\limits_{\ell,m,m'}\Im\left[\bar{f}_{(\ell,m')}\langle\ell,m'|L^{a}|\ell,m\rangle\dot{f}_{(\ell,m)}\right],\\
\langle\vec{L}\vec{L}\rangle^{ab} & \equiv\sum\limits_{\ell,m,m'}\bar{f}_{(\ell,m')}\langle\ell,m'|L^{(a}L^{b)}|\ell,m\rangle f_{(\ell,m)}.
\end{align}
\end{subequations}

The structure of Eq.~\eqref{eq:minimal rotation condition} resembles to the action
of the helical Killing vector on the waveform. Therefore the numerical
definition of $\vec{\Omega}_{\text{rot}}$ is attempting to enforce the helically
symmetric condition on the waveform modes.
Therefore it is rather circular to use this
$\vec{\Omega}_{\text{rot}}$ to analyze how well the helical symmetry is
satisfied.  Nonetheless, we perform this analysis, which agrees with
the second choice of angular velocity.

\subsubsection{Angular velocity from $(2,2)$ mode phase}
The dominant $(2,2)$ mode can be decomposed into an amplitude $A_{22}$
and phase $\phi_{22}$, which can be used to obtain the angular
velocity via%
\begin{subequations}
\begin{align}
h_{22} &= A_{22}(t) \, e^{i \, \phi_{22}(t)} \,,\\
\Omega_{22} &=  \left| \frac{1}{2}\frac{d \phi_{22}(t)}{dt} \right|
\,.
\end{align}
\end{subequations}
We compare these different choices below.

\subsubsection{Results}
\label{sec:results}

\begin{figure}
  \centering
  \includegraphics{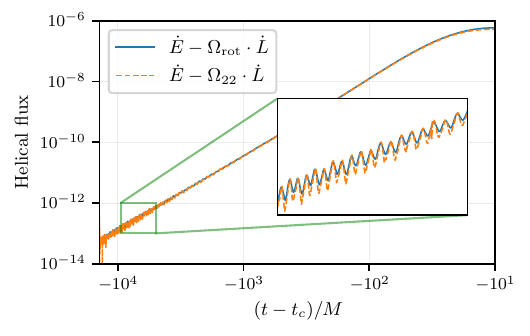}
  \caption{%
    Oscillations in the helical flux computed for a quasi-circular
    non-precessing binary (SXS:BBH:1154) with $e=5.68 \times
    10^{-5}$. Both angular velocity
    definitions---$\Omega_{\text{rot}}$ (blue) and $\Omega_{22}$
    (orange)---exhibit modulations in the helical flux.}
    \label{fig:helical_flux compare}
\end{figure}%

These definitions can be used to compute the angular velocity
numerically.
We presented the helical fluxes from an eccentric, a precessing, and
a non-precessing quasi-circular system in
Fig.~\ref{fig:ecc-prec-qc-flux}, demonstrating the cancellations at
different PN orders for the three systems.
Now we focus just on quasi-circular systems, which have the highest
degree of cancellation.

In Fig.~\ref{fig:helical_flux compare}, we present the
helical flux computed using the two definitions of angular velocity.
The modulations present in helical flux are on the orbital timescales for both
curves. This suggests that the oscillations are due to the small eccentricity
($e < 10^{-4}$, for our case) of the binary.  For a perfectly quasi-circular
binary, this flux would be smooth. Therefore we have to filter out or remove
these orbital oscillations to disentangle the effect of eccentricity. This is
achieved by performing moving averages on orbital timescales of the required
quantities. The moving average is numerically performed by convolving the
function with an appropriate kernel that filters out the high-frequency oscillations in the time domain. We use a normalized Gaussian kernel
centered at a given $t_i$ with variance given by the orbital timescale near
$t_i$. The Gaussian kernel is
\begin{equation}
G(t - t_i) =  \frac{1}{N} \exp \left(- \frac{1}{2} \frac{(t - t_i)^{2}}{\sigma_{i}^{2}} \right),
\end{equation}
where $N = \sum_{j} G(t_j - t_i) \ \Delta t_j $ is the normalization factor and
$\Delta t_j = t_j - t_{j-1}$.
The moving average is represented by $\langle \cdot \rangle$ and performed numerically as 
\begin{align}
\langle f_i \rangle =& \sum_{j} G(t_j - t_i) \ f_j \ \Delta t_j, \nonumber \\
 =& \sum_{j} \frac{1}{N} \exp \left(- \frac{1}{2} \frac{(t_j - t_i)^{2}}{\sigma_{i}^{2}} \right) f_j \ \Delta t_j,
\end{align}
where $\sigma_i = (2 \pi / \Omega(t_i))$.
We implement the following numerical procedure to get a
prediction of the instantaneous PN order from each waveform:
\begin{enumerate}
\item Obtain the longest segment of time when $|\Omega|$ is
  monotonically increasing.
\item Compute the helical flux ${F}_{\xi}$ using Eq.~\eqref{eq: helical
flux} in this monotonically increasing segment.
\item Perform a moving average to smooth the modulations in ${F}_{\xi}$.
\item Construct a cubic spline of $\log{\langle{F_{\xi}}\rangle}$ as a
  function of $\log \Omega$.
\item Compute the instantaneous PN order using spline derivatives.
\item Average out the oscillations in the instantaneous PN order
computation to get a smooth curve via
\begin{equation}
  \label{eq:smoothed-pn-order}
    \textbf{Smoothed PN Order} = \frac{3}{2} \times \bigg\langle \frac{\mathrm{d}\log \langle F \rangle}{\mathrm{d}\log  \Omega } \bigg\rangle.
\end{equation}
\end{enumerate}
All the resultant quantities are computed in the longest monotonic
segment of $|\Omega|$.
Eccentricity can make either $\Omega_{\text{rot}}$ or $\Omega_{22}$ be
oscillatory, limiting the range of times when we can construct the
spline.
\begin{figure}
    \centering
    \includegraphics{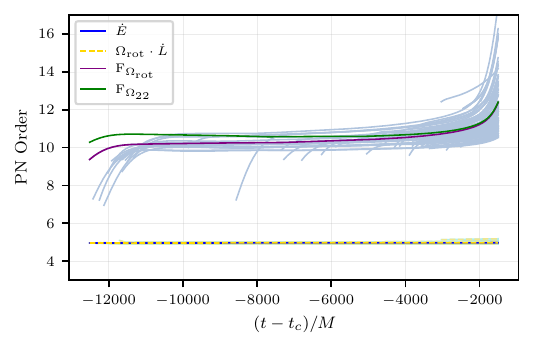}
    \caption{%
      The instantaneous PN order computed for 113 SXS simulations. The
      curves in the foreground are for a quasi-circular non-precessing
      simulation, SXS:BBH:1154.  The helical flux is computed using
      both $\Omega_{\text{rot}}$ (purple) and $\Omega_{(2,2)}$
      (green).  For reference, the yellow and blue curves correspond
      to the instantaneous PN orders of $\dot{E}$ and
      $\Omega \cdot \dot{L}$ respectively.  The curvature near the
      edges is a numerical artifact of the moving average procedure.}
    \label{fig:PN_order}
\end{figure}
We plot the smoothed PN order in Fig.~\ref{fig:PN_order}. We find that the
instantaneous PN order of the helical flux is at 5PN order relative to the
energy flux. We can clearly see that both angular velocities give comparable
predictions of the instantaneous PN order.
These numerical results support the analytical calculation presented
in Sec.~\ref{sec:abstr-comp-helic} thus establishing that the effect of
radiation reaction breaks the symmetry at such a high PN order.

\section{Discussion and conclusions}
\label{sec:disc-concl}

In this work  we investigated the approximate helical symmetry observed in
compact binary systems. The effects of eccentricity, precession, and radiation
reaction push a binary system away from being helically symmetric. We
quantify the breaking of this symmetry by computing the flux of the charge
corresponding to the symmetry. Each of these effects can be studied using
numerical waveforms by systematically controlling the initial orbital parameters
of the binary.

For quasi-circular non-precessing binary systems, we analytically
predicted that the dominant contribution to the helical flux comes
from the quadrupole moment.  The PN order of the helical flux starts
at a 5PN order relative to its contribution to the energy flux. The
analytical prediction agrees with our numerical computation of the
instantaneous PN order of the helical flux. A robust analytical
comparison requires PN results that are not yet available in the
literature. We leave computing the exact analytical expression of
helical flux to future work. Our numerical analysis justifies previous
intuition that circular orbits remains circular up to 5PN order during
the inspiral phase under radiation reaction~\cite{Apostolatos:1993nu}.
Semi-analytic waveform models, such as EOB or IMRPhenom, should
satisfy the approximate helical symmetry to a degree consistent with
NR waveforms to ensure their accuracy.
The helical flux can therefore act as a consistency check on these
waveform models.

In the numerical analysis, we realized a subtlety in the choice of
angular velocity ($\Omega$) used to compute the helical flux.
We used two definitions of $\Omega$: $\Omega_{\text{rot}}$ minimizes
the waveform's time dependence in a corotating frame, and
$\Omega_{22}$ is obtained from the phase of the dominant $(2,2)$ mode
of the strain.
The subtlety here is that $\Omega_{\text{rot}}$ effectively tries to
impose a helical symmetry, thus making the calculation of helical flux
somewhat circular.
Nonetheless, these two choices give comparable computations of
the instantaneous PN order.

We have been wary of the oscillations observed in the helical flux for
quasi-circular non-precessing binary.  The initial data for these
simulations are not perfectly quasi-circular, thus small initial
eccentricities produce modulations in the helical flux.  Therefore, we
chose to filter out these modulations by performing Gaussian
convolution.

For eccentric non-precessing binary systems, we analytically computed the
instantaneous helical flux. As expected, the helical flux is non-vanishing at
the leading 0PN order relative to the energy flux. In the limit of small
eccentricity, the helical flux has a linear relation to eccentricity. It would
be interesting to compare this relation using numerical waveforms.
However, we will need a different definition of angular velocity for eccentric
systems.  This comparison is left for future work.

For precessing binary systems, we expect moderate cancellation in the helical
flux. As observed in Fig.~\ref{fig:ecc-prec-qc-flux}, precession is the second
largest effect breaking the helical symmetry. A robust analytical computation is
required for obtaining the helical flux with PN results to sufficiently high order. The numerical analysis runs into
similar subtleties as in the case of quasi-circular non-precessing
binaries.
We will need a well-justified numerical splitting between orbital angular
velocity and precession angular velocity, which agrees with the
post-Newtonian prediction of precession.
For both precession and eccentricity, we will need a more careful
averaging procedure to avoid filtering out real physical modulations.
A careful analysis of these effects is left for future work.

\acknowledgments
We would like to thank Aaron Zimmerman, Katerina Chatziioannou, Davide
Gerosa, David Trestini, Éanna Flanagan, Alexandre Le Tiec, and Harald Pfeiffer
for valuable discussions and feedback on this work.
This work was supported by NSF CAREER Award PHY--2047382 and a Sloan
Foundation Research Fellowship.

\appendix

\section{Computation of memory intergral}
The flux computation involves hereditary integrals arising from both tail and
non-linear interactions. For example, the dominant tail starts at 1.5PN order
and the non-linear interaction first appears at 2.5PN order. These
integrals for the $l=2$ moment take the form
\begin{widetext}
  \begin{align}
    \text{U}_{ij}^{1.5\text{PN}} &= \frac{2GM}{c^3} \int_{0}^{\infty} d \tau \ \text{M}_{ij}^{(4)}(u-\tau) \left[ \ln \left(\frac{c\tau}{2b_0} \right) + \frac{11}{12} \right] \label{eq:1p5PN tail} ,\\
    \text{U}_{ij}^\text{2.5PN} &= \frac{G}{c^5} \bigg\{ - \frac{2}{7} \int_0^{\infty}  d \tau \left[ \text{M}^{(3)}_{a \langle i} \text{M}^{(3)}_{j\rangle a}\right](u-\tau)  +  \frac{1}{7} \text{M}^{(5)}_{a\langle i}\text{M}_{j\rangle a}  - \frac{5}{7} \,\text{M}^{(4)}_{a\langle i}\text{M}^{(1)}_{j\rangle a}  -\frac{2}{7}\,\text{M}^{(3)}_{a\langle i}\text{M}^{(2)}_{j\rangle a}  + \frac{1}{3}\epsilon^{}_{ab\langle i}\text{M}^{(4)}_{j\rangle a}\text{S}^{}_{b} \bigg\} \, \label{eq:2p5PN tail},  
    \end{align}
\end{widetext}
where $\text{M}_{ij}$ and $S_{i}$ are the canonical mass and spin multipole
moments, respectively, and $b_{0}$ is an arbitrary constant that cancels out at
the end of the flux calculation. We follow the procedure outlined in
\cite{Blanchet:1996wx,Blanchet:2023sbv,Blanchet:2024mnz} to compute these
integrals. By expanding canonical multipole moments using the PN series
derived in \cite{Faye:2012we,Faye:2014fra, Marchand:2016vox,Blanchet:2022vsm, Trestini:2022tot}, we obtain the
hereditary integrals in terms of the source and gauge multipole moments.
The latest PN results for source and gauge multipole moments in terms of
binary's parameters have been provided in
\cite{Marchand:2020fpt,Larrouturou:2021dma,Larrouturou:2021gqo,Henry:2021cek}. At this stage
we incorporate the quasicircular behavior of binary's orbit and use the
equations of motion to compute derivatives. For the case of quasicircular
orbits, an adiabatic approximation has been used in earlier work to compute the
integrals up to 3.5PN precision. This approximation assumes that the radius ($r$)
and orbital frequency ($\Omega$) of the binary remain constant during
integration. However, this approximation breaks for the dominant tail,
Eq.~\eqref{eq:1p5PN tail}, that is needed up to 2.5PN precision for the 4PN
flux computation. The higher order hereditary terms, like Eq.~\eqref{eq:2p5PN
tail}, can still be handled using the adiabatic approximation.

As shown in~\cite{Blanchet:1996wx}, the hereditary integrals involve many scalar
products between the orbital vectors $x^i$ or $v^i$, evaluated at the current
time $u$ or at earlier time $(u-\tau)$.
The products we need to perform the integrals are
\begin{subequations}
  \begin{align}
    x^i \, x_i^{\prime} &= r \, r^{\prime} \cos{(\phi - \phi^{\prime})} \, , \\
    x^i \, v_i^{\prime } &= r \, r^{\prime} \Omega^{\prime} \sin{(\phi - \phi^{\prime})} +  r \, \dot{r}^{\prime } \cos{(\phi - \phi^{\prime})} \, ,\\
    v^i  x_i^{\prime} &= - r^{\prime} \, r \, \Omega \sin{(\phi - \phi^{\prime})} +  r^{\prime} \, \dot{r} \cos{(\phi - \phi^{\prime})} \, ,\\
    v^i v_i^{\prime }&=  r \, \Omega \, r^{\prime} \, \Omega^{\prime} \cos{(\phi - \phi^{\prime})} + \dot{r} \, r^{\prime} \, \Omega^{\prime} \sin{(\phi - \phi^{\prime})} \nonumber \\ 
    & \quad - r \, \Omega \,  \dot{r}^{\prime } \sin{(\phi - \phi^{\prime})} + \mathcal{O}(x^{5}) \, .
  \end{align}
\end{subequations}
Here unprimed and primed quantities refer to the state of the binary
at two different times, e.g. at $u$ and at $u-\tau$.
For the integrals that require post-adiabatic corrections, we incorporate time
(or $\tau$) dependence on $r'$ and $\Omega'$. By converting these variables
involved in terms of the parameter $y = \left( \frac{G m \Omega}{c^3}
\right)^{2/3}$, these expressions reduce to the type of integrals presented in
Section 3.4.2 of \cite{Blanchet:2024mnz}. Thus we use standard formulas to
compute these hereditary integrals and truncate the result to the required PN
order. These contributions are combined with the instantaneous terms to obtain
the total flux.

\bibliographystyle{JHEP}
\bibliography{notes-biblio}

\providecommand{\href}[2]{#2}\begingroup\raggedright\begin{thebibliography}{10}

\bibitem{Gourgoulhon:2001ec}
E.~Gourgoulhon, P.~Grandclement and S.~Bonazzola, \emph{{Binary black holes in circular orbits. 1. A Global space-time approach}}, \href{https://doi.org/10.1103/PhysRevD.65.044020}{\emph{Phys. Rev. D} {\bfseries 65} (2002) 044020} [\href{https://arxiv.org/abs/gr-qc/0106015}{{\ttfamily gr-qc/0106015}}].

\bibitem{Grandclement:2001ed}
P.~Grandclement, E.~Gourgoulhon and S.~Bonazzola, \emph{{Binary black holes in circular orbits. 2. Numerical methods and first results}}, \href{https://doi.org/10.1103/PhysRevD.65.044021}{\emph{Phys. Rev. D} {\bfseries 65} (2002) 044021} [\href{https://arxiv.org/abs/gr-qc/0106016}{{\ttfamily gr-qc/0106016}}].

\bibitem{Blanchet:2023sbv}
L.~Blanchet, G.~Faye, Q.~Henry, F.~Larrouturou and D.~Trestini, \emph{{Gravitational-wave flux and quadrupole modes from quasicircular nonspinning compact binaries to the fourth post-Newtonian order}}, \href{https://doi.org/10.1103/PhysRevD.108.064041}{\emph{Phys. Rev. D} {\bfseries 108} (2023) 064041} [\href{https://arxiv.org/abs/2304.11186}{{\ttfamily 2304.11186}}].

\bibitem{Bernard:2017ktp}
L.~Bernard, L.~Blanchet, G.~Faye and T.~Marchand, \emph{{Center-of-Mass Equations of Motion and Conserved Integrals of Compact Binary Systems at the Fourth Post-Newtonian Order}}, \href{https://doi.org/10.1103/PhysRevD.97.044037}{\emph{Phys. Rev. D} {\bfseries 97} (2018) 044037} [\href{https://arxiv.org/abs/1711.00283}{{\ttfamily 1711.00283}}].

\bibitem{Blanchet:2024mnz}
L.~Blanchet, \emph{{Post-Newtonian theory for gravitational waves}}, \href{https://doi.org/10.1007/s41114-024-00050-z}{\emph{Living Rev. Rel.} {\bfseries 27} (2024) 4}.

\bibitem{Faye:2012we}
G.~Faye, S.~Marsat, L.~Blanchet and B.R.~Iyer, \emph{{The third and a half post-Newtonian gravitational wave quadrupole mode for quasi-circular inspiralling compact binaries}}, \href{https://doi.org/10.1088/0264-9381/29/17/175004}{\emph{Class. Quant. Grav.} {\bfseries 29} (2012) 175004} [\href{https://arxiv.org/abs/1204.1043}{{\ttfamily 1204.1043}}].

\bibitem{Arun:2007rg}
K.G.~Arun, L.~Blanchet, B.R.~Iyer and M.S.S.~Qusailah, \emph{{Tail effects in the 3PN gravitational wave energy flux of compact binaries in quasi-elliptical orbits}}, \href{https://doi.org/10.1103/PhysRevD.77.064034}{\emph{Phys. Rev. D} {\bfseries 77} (2008) 064034} [\href{https://arxiv.org/abs/0711.0250}{{\ttfamily 0711.0250}}].

\bibitem{Blanchet:1996wx}
L.~Blanchet, \emph{{Energy losses by gravitational radiation in inspiraling compact binaries to five halves postNewtonian order}}, \href{https://doi.org/10.1103/PhysRevD.71.129904}{\emph{Phys. Rev. D} {\bfseries 54} (1996) 1417} [\href{https://arxiv.org/abs/gr-qc/9603048}{{\ttfamily gr-qc/9603048}}].

\bibitem{LeTiec:2011ab}
A.~Le~Tiec, L.~Blanchet and B.F.~Whiting, \emph{{The First Law of Binary Black Hole Mechanics in General Relativity and Post-Newtonian Theory}}, \href{https://doi.org/10.1103/PhysRevD.85.064039}{\emph{Phys. Rev. D} {\bfseries 85} (2012) 064039} [\href{https://arxiv.org/abs/1111.5378}{{\ttfamily 1111.5378}}].

\bibitem{Ramond:2022ctc}
P.~Ramond and A.~Le~Tiec, \emph{{First law of mechanics for spinning compact binaries: Dipolar order}}, \href{https://doi.org/10.1103/PhysRevD.106.044057}{\emph{Phys. Rev. D} {\bfseries 106} (2022) 044057} [\href{https://arxiv.org/abs/2202.09345}{{\ttfamily 2202.09345}}].

\bibitem{Boyle:2019kee}
M.~Boyle et~al., \emph{{The SXS Collaboration catalog of binary black hole simulations}}, \href{https://doi.org/10.1088/1361-6382/ab34e2}{\emph{Class. Quant. Grav.} {\bfseries 36} (2019) 195006} [\href{https://arxiv.org/abs/1904.04831}{{\ttfamily 1904.04831}}].

\bibitem{Noether:1918zz}
E.~Noether, \emph{{Invariant Variation Problems}}, \href{https://doi.org/10.1080/00411457108231446}{\emph{Gott. Nachr.} {\bfseries 1918} (1918) 235} [\href{https://arxiv.org/abs/physics/0503066}{{\ttfamily physics/0503066}}].

\bibitem{Apostolatos:1993nu}
T.~Apostolatos, D.~Kennefick, E.~Poisson and A.~Ori, \emph{{Gravitational radiation from a particle in circular orbit around a black hole. 3: Stability of circular orbits under radiation reaction}}, \href{https://doi.org/10.1103/PhysRevD.47.5376}{\emph{Phys. Rev. D} {\bfseries 47} (1993) 5376}.

\bibitem{Bondi:1962px}
H.~Bondi, M.G.J.~van~der Burg and A.W.K.~Metzner, \emph{{Gravitational waves in general relativity. 7. Waves from axisymmetric isolated systems}}, \href{https://doi.org/10.1098/rspa.1962.0161}{\emph{Proc. Roy. Soc. Lond. A} {\bfseries 269} (1962) 21}.

\bibitem{Sachs:1962zza}
R.~Sachs, \emph{{Asymptotic symmetries in gravitational theory}}, \href{https://doi.org/10.1103/PhysRev.128.2851}{\emph{Phys. Rev.} {\bfseries 128} (1962) 2851}.

\bibitem{Sachs:1962wk}
R.K.~Sachs, \emph{{Gravitational waves in general relativity. 8. Waves in asymptotically flat space-times}}, \href{https://doi.org/10.1098/rspa.1962.0206}{\emph{Proc. Roy. Soc. Lond. A} {\bfseries 270} (1962) 103}.

\bibitem{Wald:1999wa}
R.M.~Wald and A.~Zoupas, \emph{{A General definition of 'conserved quantities' in general relativity and other theories of gravity}}, \href{https://doi.org/10.1103/PhysRevD.61.084027}{\emph{Phys. Rev. D} {\bfseries 61} (2000) 084027} [\href{https://arxiv.org/abs/gr-qc/9911095}{{\ttfamily gr-qc/9911095}}].

\bibitem{Flanagan:2015pxa}
E.E.~Flanagan and D.A.~Nichols, \emph{{Conserved charges of the extended Bondi-Metzner-Sachs algebra}}, \href{https://doi.org/10.1103/PhysRevD.95.044002}{\emph{Phys. Rev. D} {\bfseries 95} (2017) 044002} [\href{https://arxiv.org/abs/1510.03386}{{\ttfamily 1510.03386}}].

\bibitem{Loutrel:2016cdw}
N.~Loutrel and N.~Yunes, \emph{{Hereditary Effects in Eccentric Compact Binary Inspirals to Third Post-Newtonian Order}}, \href{https://doi.org/10.1088/1361-6382/aa59c3}{\emph{Class. Quant. Grav.} {\bfseries 34} (2017) 044003} [\href{https://arxiv.org/abs/1607.05409}{{\ttfamily 1607.05409}}].

\bibitem{LeTiec:2015kgg}
A.~Le~Tiec, \emph{{First Law of Mechanics for Compact Binaries on Eccentric Orbits}}, \href{https://doi.org/10.1103/PhysRevD.92.084021}{\emph{Phys. Rev. D} {\bfseries 92} (2015) 084021} [\href{https://arxiv.org/abs/1506.05648}{{\ttfamily 1506.05648}}].

\bibitem{Marchand:2016vox}
T.~Marchand, L.~Blanchet and G.~Faye, \emph{{Gravitational-wave tail effects to quartic non-linear order}}, \href{https://doi.org/10.1088/0264-9381/33/24/244003}{\emph{Class. Quant. Grav.} {\bfseries 33} (2016) 244003} [\href{https://arxiv.org/abs/1607.07601}{{\ttfamily 1607.07601}}].

\bibitem{Thorne:1980ru}
K.S.~Thorne, \emph{{Multipole Expansions of Gravitational Radiation}}, \href{https://doi.org/10.1103/RevModPhys.52.299}{\emph{Rev. Mod. Phys.} {\bfseries 52} (1980) 299}.

\bibitem{Faye:2014fra}
G.~Faye, L.~Blanchet and B.R.~Iyer, \emph{{Non-linear multipole interactions and gravitational-wave octupole modes for inspiralling compact binaries to third-and-a-half post-Newtonian order}}, \href{https://doi.org/10.1088/0264-9381/32/4/045016}{\emph{Class. Quant. Grav.} {\bfseries 32} (2015) 045016} [\href{https://arxiv.org/abs/1409.3546}{{\ttfamily 1409.3546}}].

\bibitem{Marchand:2020fpt}
T.~Marchand, Q.~Henry, F.~Larrouturou, S.~Marsat, G.~Faye and L.~Blanchet, \emph{{The mass quadrupole moment of compact binary systems at the fourth post-Newtonian order}}, \href{https://doi.org/10.1088/1361-6382/ab9ce1}{\emph{Class. Quant. Grav.} {\bfseries 37} (2020) 215006} [\href{https://arxiv.org/abs/2003.13672}{{\ttfamily 2003.13672}}].

\bibitem{Larrouturou:2021dma}
F.~Larrouturou, Q.~Henry, L.~Blanchet and G.~Faye, \emph{{The quadrupole moment of compact binaries to the fourth post-Newtonian order: I. Non-locality in time and infra-red divergencies}}, \href{https://doi.org/10.1088/1361-6382/ac5762}{\emph{Class. Quant. Grav.} {\bfseries 39} (2022) 115007} [\href{https://arxiv.org/abs/2110.02240}{{\ttfamily 2110.02240}}].

\bibitem{Larrouturou:2021gqo}
F.~Larrouturou, L.~Blanchet, Q.~Henry and G.~Faye, \emph{{The quadrupole moment of compact binaries to the fourth post-Newtonian order: II. Dimensional regularization and renormalization}}, \href{https://doi.org/10.1088/1361-6382/ac5ba0}{\emph{Class. Quant. Grav.} {\bfseries 39} (2022) 115008} [\href{https://arxiv.org/abs/2110.02243}{{\ttfamily 2110.02243}}].

\bibitem{Henry:2021cek}
Q.~Henry, G.~Faye and L.~Blanchet, \emph{{The current-type quadrupole moment and gravitational-wave mode (\ensuremath{\ell}, m) = (2, 1) of compact binary systems at the third post-Newtonian order}}, \href{https://doi.org/10.1088/1361-6382/ac1850}{\emph{Class. Quant. Grav.} {\bfseries 38} (2021) 185004} [\href{https://arxiv.org/abs/2105.10876}{{\ttfamily 2105.10876}}].

\bibitem{Blanchet:2022vsm}
L.~Blanchet, G.~Faye and F.~Larrouturou, \emph{{The quadrupole moment of compact binaries to the fourth post-Newtonian order: from source to canonical moment}}, \href{https://doi.org/10.1088/1361-6382/ac840c}{\emph{Class. Quant. Grav.} {\bfseries 39} (2022) 195003} [\href{https://arxiv.org/abs/2204.11293}{{\ttfamily 2204.11293}}].

\bibitem{Trestini:2022tot}
D.~Trestini, F.~Larrouturou and L.~Blanchet, \emph{{The quadrupole moment of compact binaries to the fourth post-Newtonian order: relating the harmonic and radiative metrics}}, \href{https://doi.org/10.1088/1361-6382/acb5de}{\emph{Class. Quant. Grav.} {\bfseries 40} (2023) 055006} [\href{https://arxiv.org/abs/2209.02719}{{\ttfamily 2209.02719}}].

\bibitem{JMM:xAct}
``x{A}ct: Efficient tensor computer algebra for the {W}olfram {L}anguage.'' \url{http://www.xact.es/}.

\bibitem{MARTINGARCIA2008597}
J.M.~Martín-García, \emph{x{P}erm: fast index canonicalization for tensor computer algebra}, \href{https://doi.org/10.1016/j.cpc.2008.05.009}{\emph{Comp. Phys. Commun.} {\bfseries 179} (2008) 597 } [\href{https://arxiv.org/abs/0803.0862}{{\ttfamily 0803.0862}}].

\bibitem{boyle_2024_12585016}
M.~Boyle, D.~Iozzo, L.~Stein, A.~Khairnar, H.~Rüter, M.~Scheel et~al., \emph{scri},  June, 2024.
\newblock 10.5281/zenodo.12585016.

\bibitem{Boyle:2013nka}
M.~Boyle, \emph{{Angular velocity of gravitational radiation from precessing binaries and the corotating frame}}, \href{https://doi.org/10.1103/PhysRevD.87.104006}{\emph{Phys. Rev. D} {\bfseries 87} (2013) 104006} [\href{https://arxiv.org/abs/1302.2919}{{\ttfamily 1302.2919}}].

\bibitem{OShaughnessy:2011pmr}
R.~O'Shaughnessy, B.~Vaishnav, J.~Healy, Z.~Meeks and D.~Shoemaker, \emph{{Efficient asymptotic frame selection for binary black hole spacetimes using asymptotic radiation}}, \href{https://doi.org/10.1103/PhysRevD.84.124002}{\emph{Phys. Rev. D} {\bfseries 84} (2011) 124002} [\href{https://arxiv.org/abs/1109.5224}{{\ttfamily 1109.5224}}].

\end{thebibliography}\endgroup

\end{document}